\documentclass[%
superscriptaddress,
preprint,
 amsmath,amssymb,
 aps,
prstab,
]{revtex4-2}

\usepackage{graphicx}
\usepackage{subfigure}
\usepackage{dcolumn}
\usepackage{bm}
\usepackage{multirow,ragged2e}

\begin{document}


\title{Formulae of luminosity and beam-beam tune shifts for flat-beam asymmetric colliders}

\author{Demin Zhou}
\email{dmzhou@post.kek.jp}
\affiliation{%
 KEK, 1-1 Oho, Tsukuba 305-0801, Japan 
}%
\affiliation{
 The Graduate University for Advanced Studies, SOKENDAI
}%


\date{\today}

\begin{abstract}
This note outline the formulae of luminosity and beam-beam tune shifts applicable to flat-beam asymmetric colliders. The formulae are tested with the machine parameters of SuperKEKB.
\end{abstract}

\maketitle


\section{Formulae of luminosity}\label{sec:Lum_formulations}
The luminosity of a collider can be calculated by performing the overlap integral of the 3D distributions of the colliding beams~\cite{HerrCAS2006}
\begin{equation}
    L=N_+N_-f_c K \int d^3\Vec{x} ds_0
    \rho_+(\Vec{x},-s_0) \rho_-(\Vec{x},s_0),
\end{equation}
with $f_c$ the collision frequency, $N_\pm$ the bunch populations, $\rho_\pm(\Vec{x},\pm s_0)$ the spatial distribution of the beams, and $K=\sqrt{(\Vec{v}_+-\Vec{v}_-)^2-(\Vec{v}_+\times\Vec{v}_-)^2/c^2}$ the kinematic factor. For high-energy lepton colliders, the kinematic factor can be well approximated by $K\approx 2c\cos^2\frac{\theta_c}{2}$, with $|\Vec{v}_\pm|=c$ and $\theta_c$ the full crossing angle. Gaussian distributions are often used to describe the beams
\begin{equation}
    \rho(x,y,s,s_0)=
    \frac{e^{-\frac{x^2}{2\sigma_{x}^2(s)}-\frac{y^2}{2\sigma_{y}^2(s,x)}-\frac{(s-s_0)^2}{2\sigma_{z}^2}}}{(2\pi)^{3/2}\sigma_{x}(s)\sigma_{y}(s,x)\sigma_{z}}
    \label{eq:GaussianChargeDensity}
\end{equation}
in the beams' frames. Here the transverse beam sizes $\sigma_{x,y}$ are written as functions of the longitudinal offset because of hourglass effects
\begin{equation}
    \sigma_x(s) = \sigma_x^* \sqrt{1+s^2/\beta_x^{*2}},
\end{equation}
\begin{equation}
    \sigma_y(s,x) = \sigma_{y}^*\sqrt{1+\left( s+R_{CW} x/\tan\theta_c \right)^2/\beta_y^{*2}},
\end{equation}
with $\beta_{x,y}^*$ the beta functions at the IP, $\sigma_{x,y}^*=\sqrt{\beta_{x,y}^*\epsilon_{x,y}}$ the beam sizes at IP. The parameter $R_{CW}$ is the CW ratio, with an arbitrary value of 1 for full CW and 0 for no CW. The luminosity can be written as
\begin{equation}
    L=  \frac{N_bI_{b+}I_{b-} R_{HC}}{2\pi e^2f_0 \Sigma_x^* \Sigma_y^*} =L_0 R_{HC},
    \label{eq:Lum_RHC}
\end{equation}
where $\Sigma_u^*=\sqrt{\sigma_{u+}^{*2}+\sigma_{u-}^{*2}}$ with $u=x,y$, $R_{HC}$ the geometric factor representing the effects of crossing angle, hourglass effect, and the CW. The nominal luminosity $L_0$ is defined as a function of the number of bunches $N_b$, the bunch currents $I_{b\pm}$, the transverse beam sizes at IP, and the revolution frequency $f_0$.

With $R_{CW}=0$ and flat-beam condition $\sigma_y^* \ll \sigma_x^*$, the geometric factor can be approximated by~\cite{Hirata1995PRL}
\begin{equation}
    R_{HC}\approx
    \sqrt{\frac{2}{\pi}} a e^b K_0(b),
    \label{eq:Hourglass_factor_with_crossing_angle}
\end{equation}
where $K_0(b)$ is the modified Bessel function of the second kind, and the parameters $a$ and $b$ are
\begin{equation}
    a=
    \frac{\Sigma_y^*}{\Sigma_z\Sigma_\beta^*},
\end{equation}
\begin{equation}
    b=a^2 \left( 1+ \frac{\Sigma_z^2}{\Sigma_x^{*2}}\tan^2 \frac{\theta_c}{2} \right),
\end{equation}
with the quantities of $\Sigma_\beta^*$$=$$\sqrt{\sigma_{y+}^{*2}/\beta_{y+}^{*2}+\sigma_{y-}^{*2}/\beta_{y-}^{*2}}$ and $\Sigma_z$$=$$\sqrt{\sigma_{z+}^{2}+\sigma_{z-}^{2}}$. It is noteworthy that Eq.~(\ref{eq:Hourglass_factor_with_crossing_angle}) has the same form as shown in Ref.~\cite{Hirata1995PRL}, but here the parameters $a$ and $b$ are expressed in terms of the parameters of asymmetric beams.

With $R_{CW}=1$ and a large Piwinski angle (to be explicitly defined later), the geometric factor can be approximated as
\begin{equation}
    R_{HC}^{CW} \approx
    \frac{\Sigma_x^*\Sigma_z \tan \frac{\theta_c}{2}}{\Sigma_z^2\tan^2 \frac{\theta_c}{2}+
    \sigma_{x+}^*\sigma_{x-}^*} f(d),
    \label{eq:Hourglass_factor_with_crossing_angle_and_CW}
\end{equation}
with
\begin{equation}
    f(d)=\sqrt{\pi} d\cdot e^{d^2} \text{Erfc}(d),
\end{equation}
\begin{equation}
    d= \frac{\Sigma_y^*\Sigma_x^*}{\sqrt{2}\Sigma_\beta^* \sigma_{x+}^* \sigma_{x-}^*} \sin\theta_c.
\end{equation}
Here $\text{Erfc}(d)$ represents the complementary error function. For symmetric beams, $d$ reduces to $(\beta_y^*\sin\theta_c)/\sigma_x^*$, and Eq. (\ref{eq:Hourglass_factor_with_crossing_angle_and_CW}) will have a simpler form, which can be derived from Eq.~(15) of Ref.~\cite{Dikansky2009NIMA}.

It is seen that the horizontal beta function $\beta_x^*$ does not appear explicitly in Eqs. (\ref{eq:Hourglass_factor_with_crossing_angle}) and (\ref{eq:Hourglass_factor_with_crossing_angle_and_CW}), indicating that the horizontal hourglass effect can be neglected thanks to the flat-beam condition. When $b \gg 1$, the geometric factor Eq.~(\ref{eq:Hourglass_factor_with_crossing_angle}) reduces to
\begin{equation}
    R_{HC} \approx R_C =
    \frac{1}{\sqrt{1+ \frac{\Sigma_z^2}{\Sigma_x^{*2}}\tan^2 \frac{\theta_c}{2}}},
    \label{eq:Cross_angle_factor}
\end{equation}
 utilizing the asymptotic property of the Bessel function $K_0(x)\approx e^{-x}\sqrt{\frac{\pi}{2x}}$ for large $x$. Consequently, the vertical beta function disappears, and the crossing angle alone determines the geometric factor. This implies that the condition for neglecting the vertical hourglass effect is $b\gtrsim 1$. When there is no hourglass effect in both $x$- and $y$-directions, we have exactly $R_{HC}=R_C$. Therefore, we can define the hourglass factor for luminosity as
 \begin{equation}
     R_H=R_{HC}/R_C.
 \end{equation}

From the luminosity formula Eq.~(\ref{eq:Hourglass_factor_with_crossing_angle}), we can recognize three important parameters that fundamentally define the luminosity and also the physics of beam-beam (BB) interaction in flat-beam colliders:
\begin{itemize}
    \item $\phi_{XZ}=\frac{\Sigma_z}{\Sigma_x^*}\tan\frac{\theta_c}{2}$, the ratio of projected horizontal beam size (i.e., $\Sigma_z \tan\frac{\theta_c}{2}$) to the nominal horizontal beam size at IP. For symmetric beams, it reduces to $\phi_P=\frac{\sigma_z}{\sigma_x^*}\tan\frac{\theta_c}{2}$, the well-known Piwinski angle. The quantity $\phi_{XZ}$ is an extension of $\phi_P$ to asymmetric colliders, with $\Sigma_u$ indicating the square root of the sum of the squares (SRSS) of the beam sizes in the $u$-direction ($u=x,z$).
    \item $\phi_{HC}=\frac{\Sigma_y^*}{\Sigma_\beta^*\Sigma_x^*}\tan\frac{\theta_c}{2}$, the ratio of weighted vertical beta function to the projected bunch length (i.e., $\Sigma_x^*/\tan\frac{\theta_c}{2}$). For asymmetric beams, it is formulated in terms of SRSSs. For symmetric (i.e., $\beta_{y+}^*=\beta_{y-}^*=\beta_y^*$), it reduces to $\phi_{HC}=\frac{\beta_{y}^*}{\Sigma_{x}^*}\tan\frac{\theta_c}{2}$. This parameter is important in the crab waist collision scheme where $\phi_P\gg 1$ is required. We tentatively take it as the hourglass factor for the crab-waist collision. 
    \item $\phi_{H}=a=\frac{\Sigma_y^*}{\Sigma_z\Sigma_\beta^*}$, the ratio of weighted vertical beta function to the bunch length. For symmetric beams, it reduces to $\phi_{H}=\beta_y^*/\sigma_z$, the popular parameter defining the achievable $\beta_y^*$ in colliders.
\end{itemize}
When $\phi_{P}\ll 1$, then $b\propto \phi_H^2$. Consequently, $b \gtrsim 1$ requires $\phi_H\gtrsim 1$, which is the hourglass condition for colliders with head-on or small crossing-angle collisions. When $\phi_{P} \gg 1$, $b \approx\phi_{HC}^2 \gtrsim 1$ results in the requirement of $\phi_{HC}\gtrsim 1$. This is the hourglass condition for the CW collision: With given horizontal beam sizes at the IP, $\beta_y^*$ needs to be larger than $\sigma_x^*/\tan\frac{\theta_c}{2}$. On the other hand, when $\beta_y^*$ is squeezed to achieve a certain luminosity target, the horizontal beam sizes at the IP must also be scaled down to avoid the unwanted hourglass effects.

\begin{table}[hbt]
   \centering
   \caption{Machine parameters of SuperKEKB for tests of luminosity formulae. The set of "Baseline design" refers to Refs.~\cite{Ohnishi2013PTEP,SuperKEKBTDR} (Note that in Table 1 of Ref.~\cite{Ohnishi2013PTEP}, $\epsilon_y=11.5$ pm should be $\epsilon_y=12.88$ pm, according to Ref.~\cite{SuperKEKBTDR}.), and "Phase-3" refers to Table 2 (the column of 2021) of Ref.~\cite{Ohnishi2021EPJP}. The luminosity is calculated by Eqs. (\ref{eq:Lum_RHC}) and (\ref{eq:Hourglass_factor_with_crossing_angle}). The incoherent BB tune shifts $\xi_{x,y}^i$ and $\xi_{x,y}^{ih}$  are calculated by Eq.~(\ref{eq:Incoherent_BB_tune_shift}) and by numerical integration of Eq.~(7) in Ref.~\cite{Valishev2013Fermi}, respectively.}
\begin{ruledtabular}
\begin{tabular}{ccccc}
\multirow{2}{*}{Parameters} &  \multicolumn{2}{c}{Baseline design} & \multicolumn{2}{c}{Phase-3 (2021)}\\
& LER & HER  & LER & HER\\
\hline
    $I_{b}$ (mA) & 1.44 & 1.04 & 0.673 & 0.585\\ 
    $\epsilon_x$ (nm) & 3.2 & 4.6 & 4.0 & 4.6\\ 
    $\epsilon_y$ (pm) & 8.64 & 12.88 & 52.5 & 52.5\\
    $\beta_x^*$ (mm) & 32 & 25 & 80 & 60\\
    $\beta_y^*$ (mm) & 0.27 & 0.3 & 1 & 1\\
    $\sigma_{z}$ (mm) & 6 & 5 & 4.6 & 5.1\\
    $N_b$              & \multicolumn{2}{c}{2500} &  \multicolumn{2}{c}{1174} \\
    $\xi_x^i$        & 0.0028 & 0.0012 & 0.0028 & 0.0030 \\
    $\xi_y^i$        & 0.078 & 0.074 & 0.0432 & 0.0314 \\
    $\xi_x^{ih}$        & 0.0019\footnotemark[1] & 0.0007\footnotemark[1] & 0.0028 & 0.0030 \\
    $\xi_y^{ih}$        & 0.088 & 0.078 & 0.0441 & 0.0318 \\
    $\phi_{XZ}$              & \multicolumn{2}{c}{22.0} &  \multicolumn{2}{c}{11.6} \\
    $\phi_{HC}$              & \multicolumn{2}{c}{0.8} &  \multicolumn{2}{c}{1.7} \\
    $L$ ($10^{34}\text{ cm}^{-2}\text{s}^{-1}$)    & \multicolumn{2}{c}{80.7} &  \multicolumn{2}{c}{3.0} \\
\end{tabular}
\end{ruledtabular}
\footnotetext[1]{The numerical integrals for $\xi_x^{ih}$ using \texttt{Mathematica} do not converge well for these cases. This may explain the discrepancy with the results of $\xi_x^i$.}
\label{tb:parameters:LumTest}
\end{table}

We define the specific luminosity as
\begin{equation}
    L_{sp}=\frac{L}{N_bI_{b+}I_{b-}},
\end{equation}
 which is a geometric parameter indicating the potential of a collider for generating collision events in particle detectors. Using the previous formulations, it can be expressed as
\begin{equation}
  L_{sp}= 
  \frac{L_0 R_C}{N_bI_{b+}I_{b-}} R_H R_y.
 \label{eq:Lsp}
\end{equation}
Here $R_y$ is the reduction factor from the relative vertical orbit offset $\Delta_y$ of the colliding beams at IP
\begin{equation}
    R_y=e^{-\frac{\Delta_y^2}{2(\sigma_{y+}^{*2}+\sigma_{y-}^{*2})}}.
\end{equation}
With this formulation of luminosity, one can see that $L_{sp}$ depends on the achievable beam sizes at the IP and collision conditions (such as crossing angle, orbit offset, beams' tilt angles, etc.). Considering a very large Piwinski angle $\phi_P\gg 1$, the specific luminosity is approximated by
\begin{equation}
    L_{sp} \approx
    \frac{1}{2\pi e^2f_0 \Sigma_y^* \Sigma_z \tan \frac{\theta_c}{2}}.
    \label{eq:Lsp_Simple_approx1}
\end{equation}

\begin{figure}[htb]
   \centering
    \vspace{-2mm}
   \includegraphics*[width=120mm]{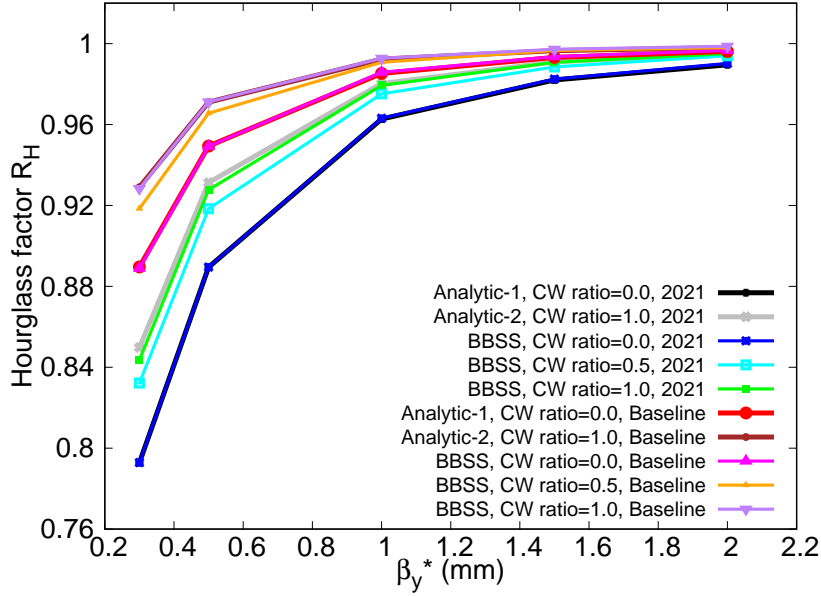}
    \vspace{-2mm}
   \caption{The hourglass factor $R_H=L/(L_0R_C)$ $=R_{HC}/R_C$ as a function of $\beta_y^*$ with beam parameters of Table~\ref{tb:parameters:LumTest}, assuming $\beta_{y+}^*=\beta_{y-}^*$. In the legends, "Analytic-1" and "Analytic-2" indicate using Eqs.~(\ref{eq:Hourglass_factor_with_crossing_angle}), (\ref{eq:Hourglass_factor_with_crossing_angle_and_CW}), and (\ref{eq:Cross_angle_factor}) to calculate $R_H$; "BBSS" indicates using BBSS code to calculate the luminosity $L$ with the CW ratio varied.}
   \label{fig:Hourglass_Factor}
   \vspace{-2mm}
\end{figure}

To check the validity of the aforementioned luminosity formulae for SuperKEKB, we use the machine parameters of Table~\ref{tb:parameters:LumTest} to perform numerical tests. The full crossing angle of SuperKEKB is $\theta_c$=83 mrad. With zero and full CW, analytic formulae are used. While using the strong-strong BB simulation code BBSS~\cite{Ohmi2000PRE, Ohmi2004PRL} to perform numerical integration with the colliding beam modeled by macroparticles, the crab waist ratio $R_{CW}$ can be arbitrarily set. In BBSS simulations, the 3D Gaussian beams are generated using optics parameters. A CW transformation is then applied to the beams before passing to the IP. Finally, the overlap integral is done to calculate the luminosity. The total charges are set to be only 1/1000 of the nominal values to avoid dynamic beam-beam effects since we are only interested in the geometric luminosity with well-defined beam distributions. The beams are tracked through only one turn, and the number of macroparticles is set to 2e6 to reduce statistic errors. The numerical results are summarized in Fig.~\ref{fig:Hourglass_Factor}. The main findings are: 1) Without the CW, Eq.~(\ref{eq:Hourglass_factor_with_crossing_angle}) is a very good approximation of luminosity for the nano-beam scheme; 2) With the CW and a large Piwinski angle, Eq.~(\ref{eq:Hourglass_factor_with_crossing_angle_and_CW}) is a fairly good approximation to the luminosity for the crab waist scheme; 3) The hourglass effect on luminosity is not negligible only when $b < 1$, considering its dependence on $\beta_y^*$ and $\sigma_x^*$; 4) The CW modifies the beam distribution, causing a luminosity gain of a few percent or less; 5) With operation conditions up to July 2022 (i.e. $\beta_y^*\geq 1$ mm), the simple formula $L_0R_C$ is fairly good to describe the luminosity of SuperKEKB. Consequently, using this formula to estimate the beam sizes at the IP from measured luminosity is also valid.

\section{Formulae of beam-beam tune shifts}

 The BB interaction will cause betatron tune shifts, which are important parameters to measure the luminosity potential of a collider. We follow the formulations of Ref.~\cite{Valishev2013Fermi} to derive the analytic formulae of incoherent BB tune shifts. The amplitude dependence of BB tune shifts is beyond the scope of this note.

 The electromagnetic fields of a three-dimensional Gaussian bunch (Here, we take it as an electron beam with the subscript ``$_-$'' indicating its fields and beam parameters) can be expressed by
\begin{subequations}
 \begin{equation}
     E_{x-}(x,y,z,t) =
     \frac{eN_-\gamma_-x}{2\epsilon_0\pi^{3/2}}
     \int_0^\infty dw
     \frac{e^{-\frac{x^2}{2\sigma_{x-}^2(s)+w}-\frac{y^2}{2\sigma_{y-}^2(s)+w}-\frac{\gamma_-^2(z-s)^2}{2\gamma_-^2\sigma_{z-}^2+w}}}
     {\left(2\sigma_{x-}^2(s)+w\right)^{3/2}\left(2\sigma_{y-}^2(s)+w\right)^{1/2}\left(2\gamma_-^2\sigma_{z-}^2+w\right)^{1/2}},
     \label{eq:Field_Ex_electron}
 \end{equation}
  \begin{equation}
     E_{y-}(x,y,z,t) =
     \frac{eN_-\gamma_-y}{2\epsilon_0\pi^{3/2}}
     \int_0^\infty dw
     \frac{e^{-\frac{x^2}{2\sigma_{x-}^2(s)+w}-\frac{y^2}{2\sigma_{y-}^2(s)+w}-\frac{\gamma_-^2(z-s)^2}{2\gamma_-^2\sigma_{z-}^2+w}}}
     {\left(2\sigma_{x-}^2(s)+w\right)^{1/2}\left(2\sigma_{y-}^2(s)+w\right)^{3/2}\left(2\gamma_-^2\sigma_{z-}^2+w\right)^{1/2}},
     \label{eq:Field_Ey_electron}
 \end{equation}
 \begin{equation}
     B_{x-}=-\frac{1}{c}E_{y-},
 \end{equation}
  \begin{equation}
     B_{y-}=\frac{1}{c}E_{x-}.
 \end{equation}
 \label{eq:BB_Fields}
 \end{subequations}
 Here $s=ct$ is taken as the orbit distance from the IP, and $t$ is interpreted as the arrival time at the IP for a particle inside the bunch. The coordinates $(x,y,z)$ give the particle's position relative to the center of the bunch. In the above expressions, the transverse beam sizes are written as a function of $s$ instead of $z$ (i.e., the beam sizes depend on the local beta functions and emittances.).

 Consider a positron bunch colliding with the electron bunch at IP with a horizontal crossing angle $\theta_c$. The centers of the two bunches arrive at the IP simultaneously at $s=0$. With $(x',y',z')$ defining a particle's coordinates in the system of the positron bunch, the coordinates transformed to the system of the electron bunch are given by
\begin{subequations}
     \begin{equation}
        x=x'\cos\theta_c + z'\sin\theta_c,
     \end{equation}
     \begin{equation}
        y=y',
     \end{equation}
     \begin{equation}
         z=z'\cos\theta_c - x'\sin\theta_c.
     \end{equation}
\end{subequations}
We are to calculate the BB tune shifts for a particle in the positron bunch with
\begin{subequations}
    \begin{equation}
        x'(t)=x_0',
    \end{equation}
    \begin{equation}
        y'(t)=y_0',
    \end{equation}
    \begin{equation}
        z'(t)=-s + z_0'.
    \end{equation}
    \label{eq:Particle_trajectory_positron_system}
\end{subequations}
Equation~(\ref{eq:Particle_trajectory_positron_system}) does not describe a particle's betatron motion around IP in a self-consistent manner. But here, we only focus on the case of $x_0'=y_0'=z_0'=0$ to calculate the incoherent BB tune shifts. In this special case, Eq.~(\ref{eq:Particle_trajectory_positron_system}) is suitable for the following calculations. Translating Eq.~(\ref{eq:Particle_trajectory_positron_system}) to the system of the electron bunch, the coordinates are
\begin{subequations}
    \begin{equation}
        x(t)=-s\sin\theta_c + x_0,
    \end{equation}
    \begin{equation}
        y(t)=y_0,
    \end{equation}
    \begin{equation}
        z(t)=-s\cos\theta_c + z_0,
    \end{equation}
    \label{eq:Particle_trajectory_electron_system}
\end{subequations}
with $x_0=x_0'\cos\theta_c+z_0'\sin\theta_c$, $y_0=y_0'$, and $z_0=z_0'\cos\theta_c-x_0'\sin\theta_c$. The Lorentz force felt by the positron particle from the electron bunch is
\begin{subequations}
    \begin{equation}
        F_{x+}(x',y',z')=e\left(E_{x-}-v_{z+}B_{y-}\right)=eE_{x-}\left(1+\cos\theta_c\right),
    \end{equation}
    \begin{equation}
        F_{y+}(x',y',z')=e\left(E_{y-}+v_{z+}B_{x-}\right)=eE_{y-}\left(1+\cos\theta_c\right),
    \end{equation}
\end{subequations}
with $E_{x-}$ and $E_{y-}$ given by Eqs.~(\ref{eq:Field_Ex_electron}) and~(\ref{eq:Field_Ey_electron}) after coordinates transformations. Integrating the BB force weighted by local $\beta$ functions yields the incoherent BB tune shifts
\begin{subequations}
    \begin{equation}
        \xi_{x+}^{ih}=\frac{1}{4\pi p_0c}\int_{-\infty}^\infty ds
        \beta_{x+}(s)\frac{\partial F_{x+}}{\partial x'},
    \end{equation}
    \begin{equation}
        \xi_{y+}^{ih}=\frac{1}{4\pi p_0c}\int_{-\infty}^\infty ds
        \beta_{y+}(s)\frac{\partial F_{y+}}{\partial y'},
    \end{equation}
    \label{eq:BB_tune_shifts_definition}
\end{subequations}
with the gradients of the BB force evaluated at $x_0'=y_0'=z_0'=0$. The hourglass effects originate from $s$-dependence of beam sizes and beta functions
\begin{subequations}
    \begin{equation}
        \beta_{u+}=\beta_{u+}^*\left(1 + \frac{s^2}{\beta_{u+}^{*2}} \right),
        \label{eq:Beta_function_around_IP}
    \end{equation}
    \begin{equation}
        \sigma_{u-}=\sigma_{u-}^*\sqrt{1 + \frac{s^2}{\beta_{u-}^{*2}} },
    \end{equation}
\end{subequations}
with $u=x,y$. Applying Eq.~(\ref{eq:BB_Fields}) to Eq.~(\ref{eq:BB_tune_shifts_definition}), the incoherent BB tune shifts can be formulated as
\begin{subequations}
 \begin{equation}
     \xi_{x+}^{ih} =
     \frac{\Lambda_+}{\sqrt{\pi}}
     \int_0^\infty dw \int_{-\infty}^\infty ds
     \frac{\gamma_-\left( 1+\cos\theta_c \right) \beta_{x+}(s) g_{x+}(s) e^{-\frac{s^2\sin^2\theta_c}{2\sigma_{x-}^2(s)+w}-\frac{\gamma_-^2s^2\left( 1+\cos\theta_c \right)^2}{2\gamma_-^2\sigma_{z-}^2+w}}}
     {\left(2\sigma_{x-}^2(s)+w\right)^{3/2}\left(2\sigma_{y-}^2(s)+w\right)^{1/2}\left(2\gamma_-^2\sigma_{z-}^2+w\right)^{1/2}},
     \label{eq:BB_tune_shifts_xix}
 \end{equation}
  \begin{equation}
     \xi_{y+}^{ih} =
     \frac{\Lambda_+}{\sqrt{\pi}}
     \int_0^\infty dw \int_{-\infty}^\infty ds
     \frac{\gamma_-\left( 1+\cos\theta_c \right) \beta_{y+}(s) e^{-\frac{s^2\sin^2\theta_c}{2\sigma_{x-}^2(s)+w}-\frac{\gamma_-^2s^2\left( 1+\cos\theta_c \right)^2}{2\gamma_-^2\sigma_{z-}^2+w}}}
     {\left(2\sigma_{x-}^2(s)+w\right)^{1/2}\left(2\sigma_{y-}^2(s)+w\right)^{3/2}\left(2\gamma_-^2\sigma_{z-}^2+w\right)^{1/2}},
     \label{eq:BB_tune_shifts_xiy}
 \end{equation}
 \label{eq:BB_tune_shifts_explicit_form}
 \end{subequations}
 with
 \begin{equation}
     \Lambda_+= \frac{r_eN_-}{2\pi\gamma_+},
 \end{equation}
 \begin{equation}
     g_{x+} (s)=
     \cos\theta_c +
     2s^2\sin\theta_c^2
     \left[
        \frac{\gamma_-^2\left( 1+\cos\theta_c \right)}{2\gamma_-^2\sigma_{z-}^2+w}
        - \frac{\cos\theta_c}{2\sigma_{x-}^2(s)+w}
     \right].
 \end{equation}
 Note that the term of $g_{x+}(s)$ in Eq.~(\ref{eq:BB_tune_shifts_xix}) is different from Eq.~(7) of Ref.~\cite{Valishev2013Fermi} as the reader can compare.

 When the hourglass effects are negligible, the $\beta$ functions and beam sizes are independent of $s$, i.e., $\beta_{u\pm}(s)=\beta_{u\pm}^*$ and $\sigma_{u\pm}(s)=\sigma_{u\pm}^*$ are constants with $u=x,y$. Consequently, the integration over $s$ in Eq.~(\ref{eq:BB_tune_shifts_explicit_form}) can be done without any approximations, yielding
 \begin{subequations}
 \begin{equation}
     \xi_{x+}^{i} =
     \Lambda_+ \beta_{x+}^*
     \int_0^\infty dw
     \frac{1}{\left(2\sigma_{y-}^{*2}+w\right)^{1/2} \left(2\overline{\sigma}_{x-}^2+\alpha_+w\right)^{3/2}},
     \label{eq:BB_tune_shifts_no_hourglass_xix1}
 \end{equation}
  \begin{equation}
     \xi_{y+}^{i} =
     \Lambda_+ \beta_{y+}^*
     \int_0^\infty dw
     \frac{1}{\left(2\sigma_{y-}^{*2}+w\right)^{3/2} \left(2\overline{\sigma}_{x-}^2+\alpha_+w\right)^{1/2}},
     \label{eq:BB_tune_shifts_no_hourglass_xiy1}
 \end{equation}
 \label{eq:BB_tune_shifts_no_hourglass1}
 \end{subequations}
 with
 \begin{equation}
     \alpha_+=1 + \frac{1}{\gamma_+^2}\tan^2\frac{\theta_c}{2},
 \end{equation}
 \begin{equation}
     \overline{\sigma}_{x-}=\sqrt{\sigma_{x-}^{*2}+\sigma_{z-}^2\tan^2\frac{\theta_c}{2}}.
 \end{equation}
 The integration over $w$ in Eq.~(\ref{eq:BB_tune_shifts_no_hourglass1}) can also be done analytically, giving
\begin{subequations}
 \begin{equation}
     \xi_{x+}^{i} =
     \frac{\Lambda_+ \beta_{x+}^*}{\sigma_{x-}^{*2}\sqrt{1+\phi_-^2}\sqrt{\alpha_+}\left(\sqrt{1+\phi_-^2}+\kappa_-\sqrt{\alpha_+}\right)},
     \label{eq:BB_tune_shifts_no_hourglass_xix2}
 \end{equation}
  \begin{equation}
     \xi_{y+}^{i} =
     \frac{\Lambda_+ \beta_{y+}^*}{\sigma_{x-}^*\sigma_{y-}^*\left(\sqrt{1+\phi_-^2}+\kappa_-\sqrt{\alpha_+}\right)},
     \label{eq:BB_tune_shifts_no_hourglass_xiy2}
 \end{equation}
 \label{eq:BB_tune_shifts_no_hourglass2}
 \end{subequations}
 with
 \begin{equation}
     \phi_- = \frac{\sigma_{z-}}{\sigma_{x-}^*}\tan\frac{\theta_c}{2},
 \end{equation}
 \begin{equation}
     \kappa_- = \frac{\sigma_{y-}^*}{\sigma_{x-}^*}.
 \end{equation}
 Here $\phi_-$ is the Piwinski angle of the electron beam, and $\kappa_-$ is the aspect ratio of the transverse beam sizes. Equation~(\ref{eq:BB_tune_shifts_no_hourglass2}) represents the solutions of the BB tune shifts for 3D Gaussian colliding bunches with an arbitrary horizontal crossing angle and beam energies but without hourglass effects. The formulae are obtained starting from Eq.~(\ref{eq:BB_Fields}) without any approximations.
 
For ultrarelativistic beams, there is $\alpha_{\pm}\rightarrow 1$, and Eq.~(\ref{eq:BB_tune_shifts_no_hourglass2}) can be written in a compact form~\cite{Raimondi2003}
 \begin{equation}
    \xi_{u\pm}^i= \frac{r_e}{2\pi\gamma_\pm} \frac{N_\mp\beta_{u\pm}^*}{\overline{\sigma}_{u\mp}(\overline{\sigma}_{x\mp}+\overline{\sigma}_{y\mp})},
    \label{eq:Incoherent_BB_tune_shift}
\end{equation}
with $u=x,y$. The beam sizes in the above equation are defined as $\overline{\sigma}_{y\pm}=\sigma_{y\pm}^*$, and $\overline{\sigma}_{x\pm}=\sqrt{\sigma_{z\pm}^2\tan^2\frac{\theta_c}{2}+\sigma_{x\pm}^{*2}}$. The formula is the same as that for a head-on collision, except that the horizontal beam size is modified, considering the projection from the longitudinal direction. The incoherent BB tune shifts depend on the opposite beam's bunch current and beam sizes.

For flat beams, there is $\kappa_\pm \ll 1$, and Eq.~(\ref{eq:Incoherent_BB_tune_shift}) can be approximated by
 \begin{equation}
    \xi_{u\pm}^i \approx \frac{r_e}{2\pi\gamma_\pm} \frac{N_\mp\beta_{u\pm}^*}{\overline{\sigma}_{u\mp}\overline{\sigma}_{x\mp}}.
    \label{eq:Incoherent_BB_tune_shift_approx1}
\end{equation}
This formula is obtained by dropping the negligible terms in Eq.~(\ref{eq:BB_tune_shifts_no_hourglass1}) as follows
 \begin{subequations}
 \begin{equation}
     \xi_{x+}^{i} \approx
     \Lambda_+ \beta_{x+}^*
     \int_0^\infty dw
     \frac{1}{w^{1/2} \left(2\overline{\sigma}_{x-}^2+w\right)^{3/2}},
     \label{eq:BB_tune_shifts_no_hourglass_xix3}
 \end{equation}
  \begin{equation}
     \xi_{y+}^{i} \approx
     \Lambda_+ \beta_{y+}^*
     \int_0^\infty dw
     \frac{1}{\left(2\sigma_{y-}^{*2}+w\right)^{3/2} \left(2\overline{\sigma}_{x-}^2\right)^{1/2}}.
     \label{eq:BB_tune_shifts_no_hourglass_xiy3}
 \end{equation}
 \label{eq:BB_tune_shifts_no_hourglass3}
 \end{subequations}
 We will use such approximations later when considering the hourglass effects ($s$-dependence of $\beta$ functions and beam sizes.).

With the hourglass effect taken into account, the BB tune shift of on-axis particles (i.e. $\xi_{u\pm}^{ih}$) can be numerically calculated by integrating the $\beta$-function weighted BB force along their path. For example, one can start from Eq.~(\ref{eq:BB_tune_shifts_explicit_form}) (i.e., Eq. (7) of Ref.~\cite{Valishev2013Fermi} with typos fixed) to perform the numerical integration. We can define the hourglass factor for BB tune shifts as
\begin{equation}
    R_{\xi u\pm}=\xi_{u\pm}^{ih}/\xi_{u\pm}^i.
\end{equation}

With approximations of ultrarelativistic beams ($\alpha_{\pm}\rightarrow 1$) and flat beams ($\kappa_\pm \ll 1$), we can also find approximate formulae for $\xi_{u\pm}^{ih}$. Here we also assume $\beta_{x\pm}^*\gg \sigma_{z\pm}$ (This is fairly satisfied in flat-beam colliders like SuperKEKB.), i.e., the hourglass effect in the horizontal direction is negligible. With all these assumptions, Eq.~(\ref{eq:BB_tune_shifts_explicit_form}) is simplified as
\begin{subequations}
 \begin{equation}
     \xi_{x+}^{ih} \approx
     \frac{\Lambda_+}{\sqrt{\pi}}
     \int_0^\infty dw \int_{-\infty}^\infty ds
     \frac{\left( 1+\cos\theta_c \right) \beta_{x+}^* g_{x+}(s) e^{-\frac{s^2\sin^2\theta_c}{2\sigma_{x-}^{*2}+w}-\frac{s^2\left( 1+\cos\theta_c \right)^2}{2\sigma_{z-}^2}}}{\left(2\sigma_{x-}^{*2}+w\right)^{3/2}\left(2\sigma_{y-}^2(s)+w\right)^{1/2}\left(2\sigma_{z-}^2\right)^{1/2}},
     \label{eq:BB_tune_shifts_xix_approx1}
 \end{equation}
  \begin{equation}
     \xi_{y+}^{ih} \approx
     \frac{\Lambda_+}{\sqrt{\pi}}
     \int_0^\infty dw \int_{-\infty}^\infty ds
     \frac{\left( 1+\cos\theta_c \right) \beta_{y+}(s) e^{-\frac{s^2\sin^2\theta_c}{2\sigma_{x-}^{*2}+w}-\frac{s^2\left( 1+\cos\theta_c \right)^2}{2\sigma_{z-}^2}}}{\left(2\sigma_{x-}^{*2}+w\right)^{1/2}\left(2\sigma_{y-}^2(s)+w\right)^{3/2}\left(2\sigma_{z-}^2\right)^{1/2}},
     \label{eq:BB_tune_shifts_xiy_approx1}
 \end{equation}
 \label{eq:BB_tune_shifts_explicit_form_approx1}
 \end{subequations}
 We only keep the $s$-dependence of vertical beam size and the beta function around IP in the above formulation. Obtaining analytic solutions for Eq.~(\ref{eq:BB_tune_shifts_explicit_form_approx1}) is not trivial and requires further approximations. For flat beams and the vertical beam size does not vary fast in the overlap region, Eq.~(\ref{eq:BB_tune_shifts_xix_approx1}) reduces to 
  \begin{equation}
     \xi_{x+}^{ih} \approx
     \frac{\Lambda_+}{\sqrt{\pi}}
     \int_0^\infty dw \int_{-\infty}^\infty ds
     \frac{\left( 1+\cos\theta_c \right) \beta_{x+}^* g_{x+}(s) e^{-\frac{s^2\sin^2\theta_c}{2\sigma_{x-}^{*2}+w}-\frac{s^2\left( 1+\cos\theta_c \right)^2}{2\sigma_{z-}^2}}}{\left(2\sigma_{x-}^{*2}+w\right)^{3/2}w^{1/2}\left(2\sigma_{z-}^2\right)^{1/2}}.
     \label{eq:BB_tune_shifts_xix_approx2}
 \end{equation}
 Indeed, the hourglass effects in horizontal and vertical directions are neglected here. Then, the integrations over $w$ and $s$ are straightforward, giving
  \begin{equation}
    \xi_{x+}^i \approx \frac{r_e}{2\pi\gamma_+} \frac{N_-\beta_{x+}^*}{\overline{\sigma}_{x-}^2}.
    \label{eq:BB_tune_shifts_xix_approx3}
\end{equation}
This expression is consistent with Eq.~(\ref{eq:Incoherent_BB_tune_shift_approx1}).

To proceed further, in Eq.~(\ref{eq:BB_tune_shifts_explicit_form_approx1}) we can change the variable $s$ by
\begin{equation}
    s'=s\sqrt{\frac{\sin^2\theta_c}{2\sigma_{x-}^{*2}+w}+\frac{\left( 1+\cos\theta_c \right)^2}{2\sigma_{z-}^2}}
    =s\sqrt{\frac{4\cos^4\frac{\theta_c}{2}\left( 2\sigma_{x-}^{*2} + 2\sigma_{z-}^2\tan^2 \frac{\theta_c}{2} +w \right)}{2\sigma_{z-}^{2} \left( 2\sigma_{x-}^{*2} +w \right)}}.
    \label{eq:Variable_substitution1}
\end{equation}
With this substitution, Eq.~(\ref{eq:BB_tune_shifts_xiy_approx1}) becomes
  \begin{equation}
     \xi_{y+}^{ih} \approx
     \frac{\Lambda_+}{\sqrt{\pi}}
     \int_0^\infty dw \int_{-\infty}^\infty ds'
     \frac{\beta_{y+}(s') e^{-s'^2}}{\left(2\overline{\sigma}_{x-}^{2}+w\right)^{1/2}\left(2\sigma_{y-}^2(s')+w\right)^{3/2}}.
     \label{eq:BB_tune_shifts_xiy_approx2}
 \end{equation}
This integral is still complicated because $s'$ contains the variable $w$. We can intentionally drop $w$ in Eq.~(\ref{eq:Variable_substitution1}) and make $s'$ simply linear to $s$:
\begin{equation}
    s'\approx
    s\sqrt{\frac{4\cos^4\frac{\theta_c}{2}\left( 2\sigma_{x-}^{*2} + 2\sigma_{z-}^2\tan^2 \frac{\theta_c}{2} \right)}{2\sigma_{z-}^{2} \cdot 2\sigma_{x-}^{*2} }}
    =s\frac{\sqrt{2}\overline{\sigma}_{x-}\cos^2\frac{\theta_c}{2}}{\sigma_{z-}\sigma_{x-}^*}.
    \label{eq:Variable_substitution2}
\end{equation}
We also drop one $w$ of Eq.~(\ref{eq:BB_tune_shifts_xiy_approx2}) as follows
  \begin{equation}
     \xi_{y+}^{ih} \approx
     \frac{\Lambda_+}{\sqrt{\pi}}
     \int_0^\infty dw \int_{-\infty}^\infty ds'
     \frac{\beta_{y+}(s') e^{-s'^2}}{\left(2\overline{\sigma}_{x-}^{2}\right)^{1/2}\left(2\sigma_{y-}^2(s')+w\right)^{3/2}},
     \label{eq:BB_tune_shifts_xiy_approx3}
 \end{equation}
 considering flat-beam condition $\overline{\sigma}_{x-}\gg \sigma_{y-}(s')$ is satisfied in the overlap region of the colliding beams. The integration over $w$ can be done first in Eq.~(\ref{eq:BB_tune_shifts_xiy_approx3}), giving
   \begin{equation}
     \xi_{y+}^{ih} \approx
     \frac{\Lambda_+}{\sqrt{\pi}}
     \int_{-\infty}^\infty ds'
     \frac{\beta_{y+}(s') e^{-s'^2}}{\left(2\overline{\sigma}_{x-}^{2}\right)^{1/2}}
     \frac{2}{\sqrt{2\sigma_{y-}^2(s')}}
     =\frac{\Lambda_+}{\sqrt{\pi}\overline{\sigma}_{x-}}
     \int_{-\infty}^\infty ds'
     \frac{\beta_{y+}(s') e^{-s'^2}}{\sigma_{y-}(s')}
     .
     \label{eq:BB_tune_shifts_xiy_approx4}
 \end{equation}
 The integration over $s'$ can be explicitly written as
 \begin{equation}
     \int_{-\infty}^\infty ds'
     \frac{\beta_{y+}(s') e^{-s'^2}}{\sigma_{y-}(s')}
     =
     \frac{\beta_{y+}^*}{\sigma_{y-}^*}
     \int_{-\infty}^\infty ds'
     \frac{e^{-s'^2}}{\sqrt{1+\frac{s'^2}{2r_-^2}}}
     +
     \frac{\beta_{y-}^{*2}}{2\beta_{y+}^*\sigma_{y-}^*r_-^2}
     \int_{-\infty}^\infty ds'
     \frac{s'^2 e^{-s'^2}}{\sqrt{1+\frac{s'^2}{2r_-^2}}},
 \end{equation}
 with
 \begin{equation}
     r_- = \frac{\beta_{y-}^*\overline{\sigma}_{x-}\cos^2\frac{\theta_c}{2}}{\sigma_{z-}\sigma_{x-}^*}.
 \end{equation}
 Note that $r_-$ reduces to $\sqrt{2}\phi_{HC}$ with conditions of $\theta_c\ll 1$, large Piwinski angle, and symmetric colliding beams. In the end, we can obtain the explicit solution for Eq.~(\ref{eq:BB_tune_shifts_xiy_approx4}) as follows
    \begin{equation}
     \xi_{y+}^{ih} \approx
     \frac{r_eN_-}{2\pi\gamma_+}
     \frac{\beta_{y+}^*}{\overline{\sigma}_{x-}\sigma_{y-}^*}
     \left[
     \sqrt{\frac{2}{\pi}} r_- e^{r_-^2} K_0(r_-^2)
     +
     \frac{\beta_{y-}^{*2}}{2\sqrt{2}\beta_{y+}^{*2}r_-}
     U\left(\frac{1}{2},0,2r_-^2\right)
     \right]
     ,
     \label{eq:BB_tune_shifts_xiy_approx5}
 \end{equation}
where $K_0(x)$ is the modified Bessel function of the second kind, and $U(a,b,z)$ is Tricomi confluent hypergeometric function. The first term in the square brackets only contains the parameters of the electron beam. It means that it results from the $s$-dependece of the charge density of the electron beam. The second term contains $\beta_{y+}^*$. This means that it results from the $s$-dependence of the $\beta$ function of the positron beam.

In the case of $r_-\gg 1$, the asymptotic approximation of Eq.~(\ref{eq:BB_tune_shifts_xiy_approx5}) is
\begin{equation}
    \xi_{y+}^{ih} \approx
     \frac{r_eN_-}{2\pi\gamma_+}
     \frac{\beta_{y+}^*}{\overline{\sigma}_{x-}\sigma_{y-}^*}
     \left(
     1
     +
     \frac{\beta_{y-}^{*2}}{4\beta_{y+}^{*2}r_-^2}
     \right)
     .
     \label{eq:BB_tune_shifts_xiy_approx6}
\end{equation}
In the limit of $r_-\rightarrow \infty$, Eq.~(\ref{eq:BB_tune_shifts_xiy_approx6}) reduces to Eq.~(\ref{eq:Incoherent_BB_tune_shift_approx1}).

\section{Discussions}

\begin{figure}[htb]
   \centering
    \vspace{-2mm}
   \includegraphics*[width=120mm]{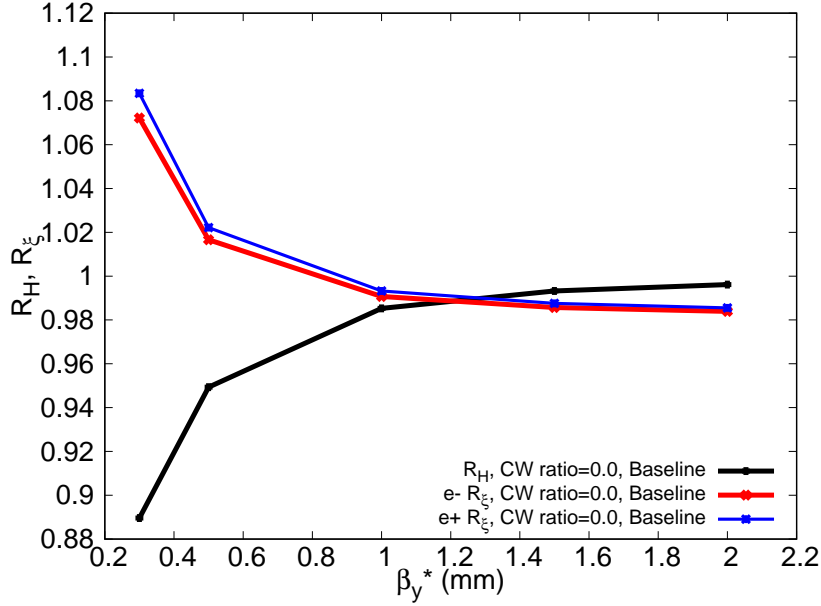}
    \vspace{-2mm}
   \caption{The hourglass factor for vertical BB tune shifts $R_{\xi y\pm}$ (blue and red lines) and luminosity $R_H$ (black line) as a function of $\beta_y^*$ with baseline design parameters of Table~\ref{tb:parameters:LumTest}, assuming $\beta_{y+}^*=\beta_{y-}^*$. $R_H=R_{HC}/R_C$ is calculated using Eqs.~(\ref{eq:Hourglass_factor_with_crossing_angle}) and (\ref{eq:Cross_angle_factor}).}
   \label{fig:Hourglass_Factor_BB_Baseline}
   \vspace{-2mm}
\end{figure}

\begin{figure}[htb]
   \centering
    \vspace{-2mm}
   \includegraphics*[width=120mm]{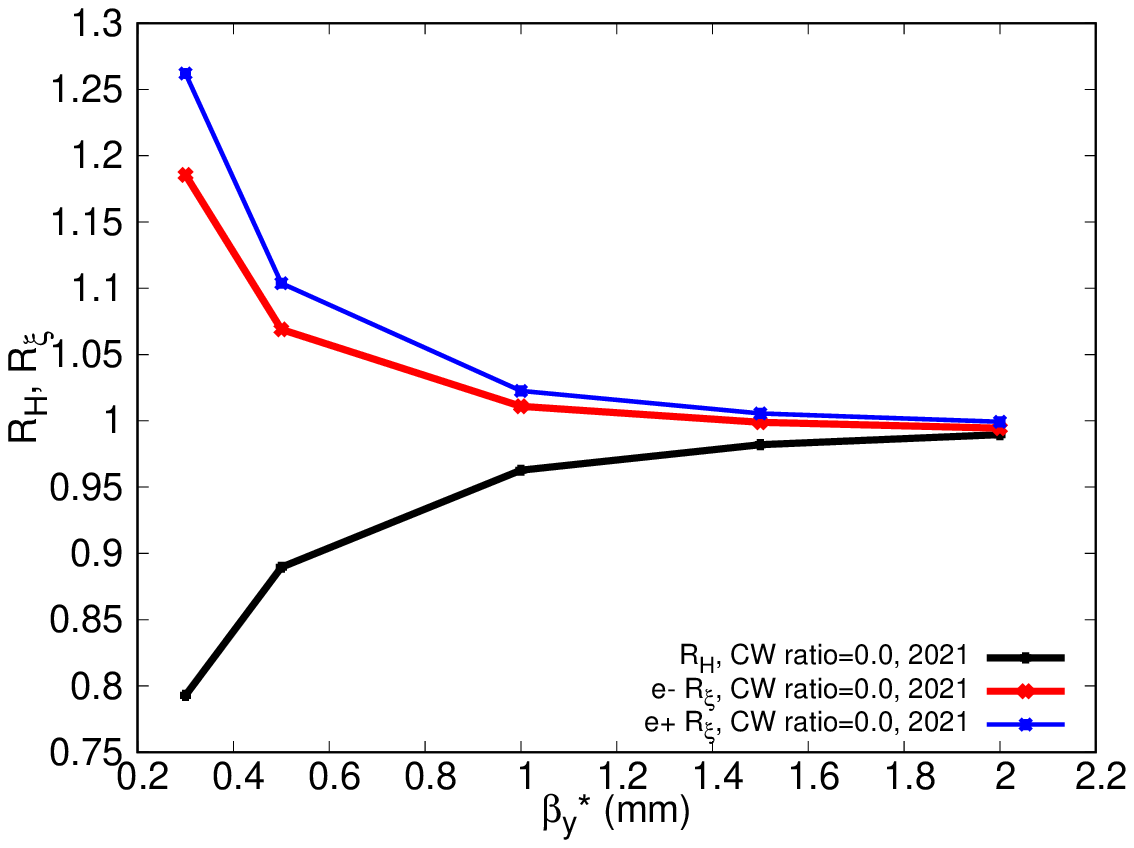}
    \vspace{-2mm}
   \caption{The hourglass factor for vertical BB tune shifts $R_{\xi y\pm}$ (blue and red lines) and luminosity $R_H$ (black line) as a function of $\beta_y^*$ with Phase-3 (2021) parameters of Table~\ref{tb:parameters:LumTest}, assuming $\beta_{y+}^*=\beta_{y-}^*$. $R_H=R_{HC}/R_C$ is calculated using Eqs.~(\ref{eq:Hourglass_factor_with_crossing_angle}) and (\ref{eq:Cross_angle_factor}).}
   \label{fig:Hourglass_Factor_BB_2021}
   \vspace{-2mm}
\end{figure}

The hourglass effect simultaneously modifies the local charge density as Eq.~(\ref{eq:GaussianChargeDensity}) and the local $\beta$ functions as Eq.~(\ref{eq:Beta_function_around_IP}). Since the charge density at $|s|>0$ is always smaller than that at $s=0$, the hourglass effect on luminosity results in the reduction factor $R_H<1$. This can be verified using the formulations of $R_{HC}$ and $R_C$ as demonstrated in Fig.~\ref{fig:Hourglass_Factor}. The combination of two factors determines the hourglass factor for the BB tune shifts: the charge density decreases, and the beta function increases as a function of $|s|$. Consequently, the quantities $R_{\xi u\pm}$ can be larger or smaller than 1. In Tables~\ref{tb:parameters:LumTest} and \ref{tb:parameters}, the incoherent BB tune shifts, i.e., $\xi_{x,y}^i$ by the simple estimate of Eq.~(\ref{eq:Incoherent_BB_tune_shift}) and $\xi_{x,y}^{ih}$ by numerical integration of Eq.~(\ref{eq:BB_tune_shifts_explicit_form}) using \texttt{Mathematica}, are compared. One can see that the two methods give results close to each other except in the case of ``Baseline design''. As a further demonstration, numerical results of hourglass factors $R_H$ (for luminosity) and $R_{\xi y}$ (for incoherent BB tune shifts) using the beam parameters of Table~\ref{tb:parameters:LumTest} are compared in Figs.~\ref{fig:Hourglass_Factor_BB_Baseline} and \ref{fig:Hourglass_Factor_BB_2021}. In this comparison, the vertical $\beta$ function at the IP is varied to reflect the $\beta_y^*$-dependence of the hourglass effect (see Eq.~(\ref{eq:Beta_function_around_IP})). It can be seen that the hourglass factors for luminosity and BB tune shift respectively decrease and increase when $\beta_y^*$ decreases (Note that smaller $\beta_y^*$ means a stronger hourglass effect). Comparing Figs.~\ref{fig:Hourglass_Factor_BB_Baseline} and \ref{fig:Hourglass_Factor_BB_2021} also show that the parameter $\beta_x^*$ strongly affects the hourglass effect as reflected from the parameter $\phi_{HC}$. It implies that squeezing $\beta_y^*$ should be performed in parallel with squeezing $\beta_x^*$ if we want to avoid strong hourglass effects.

To study coherent BB instabilities, it is more suitable to define the coherent BB tune shifts~\cite{Hirata1990NIMA}
 \begin{equation}
    \xi_{u\pm}^c= \frac{r_e}{2\pi\gamma_\pm} \frac{N_\mp\beta_{u\pm}^*}{\overline{\Sigma}_{u}(\overline{\Sigma}_{x}+\overline{\Sigma}_{y})},
    \label{eq:Coherent_BB_tune_shift}
\end{equation}
with $u=x,y$, and the effective beam size $\overline{\Sigma}_u=$ $\sqrt{\overline{\sigma}_{u+}^2+\overline{\sigma}_{u-}^2}$. Empirically, we often calculate the vertical BB parameter of flat beams from luminosity~\cite{Ohmi2004PRSTAB}
\begin{equation}
    L = \frac{1}{2e r_e} \frac{\gamma_\pm I_\pm}{\beta_{y\pm}^*} \xi_{y\pm}^L
      \simeq \frac{1}{2e r_e} \frac{\gamma_\pm I_\pm}{\beta_{y\pm}^*} \left( 2\xi_{u\pm}^c R_H \right),
\end{equation}
with $I_\pm$ the total beam currents. Here the hourglass effects are resolved in the BB parameter $\xi_{y\pm}^L$. For flat beams $\sigma_y^*\ll \sigma_x^*$, it is easy to verify that $\xi_{y\pm}^L\simeq 2\xi_{u\pm}^c R_H$. In terms of incoherent BB tune shifts with flat beams, the luminosity is expressed as
\begin{equation}
    L = \frac{1}{2e r_e} \frac{\gamma_\pm I_\pm}{\beta_{y\pm}^*} \xi_{y\pm}^i
      \frac{2\sigma_{y\pm}^*\overline{\sigma}_{x\pm}}{\Sigma_y\overline{\Sigma}_x} R_H,
\end{equation}
Finally, one can see that, for 3D Gaussian beams with identical sizes (i.e. $\sigma_{u+}^*=\sigma_{u-}^*$), there is $\xi_{y\pm}^L=\xi_{y\pm}^i R_H = 2\xi_{y\pm}^c R_H$. When the hourglass effect is negligible, the relation is further simplified: $\xi_{y\pm}^L=\xi_{y\pm}^i = 2\xi_{y\pm}^c$. Further correlation to the incoherent BB tune shifts with the hourglass effect is
\begin{equation}
    \xi_{y\pm}^{ih} =\xi_{y\pm}^{i} R_{\xi y\pm}=\xi_{y\pm}^{L} R_{\xi y\pm}/R_H.
    \label{eq:BB_paramemter_with_Hourglass_effect1}
\end{equation}
Here $\xi_{y\pm}^{ih}$ is consistent with the definition of $\xi_y$ in Eq.~(2.3) of Ref.~\cite{KEKB_Design_Report}, with the condition that the beam sizes of the two beams are equal.

Consider a very large Piwinski angle and assume that the hourglass effect is negligible; from Eq.~(\ref{eq:Incoherent_BB_tune_shift}), the incoherent vertical BB tune shift can be simplified to
\begin{equation}
    \xi_{y\pm}^i \approx
    \frac{r_e}{2\pi ef_0 \gamma_\pm \tan \frac{\theta_c}{2} }
    \frac{I_{b\mp}\beta_{y\pm}^*}{\sigma_{y\mp}^* \sigma_{z\mp}}.
    \label{eq:Incoherent_BB_tune_shift_xiy_approx1}
\end{equation}
Furthermore, we assume that a balanced collision is achievable: $\beta_{y+}^*=\beta_{y-}^*=\beta_y^*$ and $\epsilon_{y+}=\epsilon_{y-}=\epsilon_y$. The above equation can then be rewritten as
\begin{equation}
    \xi_{y\pm}^i \approx
    \frac{r_e}{2\pi ef_0 \gamma_\pm \tan \frac{\theta_c}{2} }
    \frac{I_{b\mp}}{\sigma_{z\mp}}
    \sqrt{\frac{\beta_y^*}{\epsilon_y}}.
    \label{eq:Incoherent_BB_tune_shift_xiy_approx2}
\end{equation}
If there is an upper limit on the achievable BB tune shift, the above equation shows a constraint between the bunch current, the vertical emittance, and the vertical beta function at the IP. To achieve the same BB tune shift at a given bunch current, squeezing $\beta_y^*$ requires reducing the single-beam emittance $\epsilon_y$. With fixed $\beta_y^*$, when increasing the bunch currents, we expect emittance blowup as scaled by $\epsilon_y \propto I_{b\pm}^2$.

\begin{table}[hbt]
   \centering
   \caption{SuperKEKB machine parameters for $\beta_y^*$=2 mm on Jul. 1, 2019 and $\beta_y^*$=1 mm on Apr. 5, 2022, respectively. The luminosity is calculated by Eqs. (\ref{eq:Lum_RHC}) and (\ref{eq:Hourglass_factor_with_crossing_angle}). The incoherent BB tune shifts $\xi_{x,y}^i$ and $\xi_{x,y}^{ih}$  are calculated by Eq.~(\ref{eq:Incoherent_BB_tune_shift}) and by numerical integration of Eq.~(7) in Ref.~\cite{Valishev2013Fermi}, respectively.}
\begin{ruledtabular}
\begin{tabular}{ccccc}
\multirow{2}{*}{Parameters} &  \multicolumn{2}{c}{2019.07.01} & \multicolumn{2}{c}{2022.04.05}\\
& LER & HER  & LER & HER\\
\hline
    $I_{b}$ (mA) & 0.51 & 0.51 & 0.71 & 0.57\\ 
    $\epsilon_x$ (nm) & 2.0 & 4.6 & 4.0 & 4.6\\ 
    $\epsilon_y$ (pm) & 40 & 40 & 30 & 35\\
    $\beta_x$ (mm) & 80 & 80 & 80 & 60\\
    $\beta_y$ (mm) & 2 & 2 & 1 & 1\\
    $\sigma_{z0}$ (mm) & 4.6 & 5.0 & 4.6 & 5.1\\
    $\nu_{x}$ & 44.542 & 45.53 & 44.524 & 45.532\\
    $\nu_{y}$ & 46.605 & 43.583 & 46.589 & 43.572\\
    $\nu_{s}$ & 0.023 & 0.027 & 0.023 & 0.027\\
    Crab waist ratio & 0 & 0 & 80\% & 40\% \\
    $N_b$              & \multicolumn{2}{c}{1174} &  \multicolumn{2}{c}{1174} \\
    $\xi_x^i$        & 0.0034 & 0.0023 & 0.0036 & 0.0024 \\
    $\xi_y^i$        & 0.0621 & 0.0386 & 0.0516 & 0.0438 \\
    $\xi_x^{ih}$        & 0.0034 & 0.0023 & 0.0036 & 0.0024 \\
    $\xi_y^{ih}$        & 0.0621 & 0.0383 & 0.0523 & 0.0446 \\
    $\phi_{XZ}$              & \multicolumn{2}{c}{12.3} &  \multicolumn{2}{c}{11.7} \\
    $\phi_{HC}$              & \multicolumn{2}{c}{3.6} &  \multicolumn{2}{c}{1.7} \\
    $L$ ($10^{34}\text{ cm}^{-2}\text{s}^{-1}$)    & \multicolumn{2}{c}{1.7} &  \multicolumn{2}{c}{3.9} \\
\end{tabular}
\end{ruledtabular}
\label{tb:parameters}
\end{table}

In addition to Table~\ref{tb:parameters:LumTest}, Table~\ref{tb:parameters} shows the typical machine parameters from the operation without the CW (2019.07.01) and with the CW (2022.04.05). From these parameter sets, one can see that the hourglass effect modifies the vertical incoherent tune shifts $\xi_y$ by about 11\% and 5\% respectively for LER and HER (see the difference between $\xi_y^i$ and $\xi_y^{ih}$) with the baseline design configuration of SuperKEKB. For other cases of $\beta_y^*\geq 1$ mm, the hourglass effect on the vertical incoherent tune shifts is negligible (Suppose $\beta_x^*$ satisfies $\phi_{HC}\gtrsim 1$. Consequently, Eq.~(\ref{eq:Incoherent_BB_tune_shift}) is a good approximation for BB tune shifts. The horizontal incoherent tune shifts are smaller than the vertical ones by one order with flat beams (i.e., the aspect ratios of transverse beam sizes (i.e., $\alpha_{\pm}$) and $\beta$ functions are small).

\begin{acknowledgements}
The author thanks K. Oide, M. Zobov, and A.V. Bogomyagkov for many discussions.
\end{acknowledgements}

\nocite{*}

\bibliography{aps_bb_skb}

\end{document}